\documentclass[prb,twocolumn,showpacs,preprintnumbers,amsmath,amssymb]{revtex4}

\usepackage{epsfig}

\begin{document}

\preprint{
\hfill  TH01.1
} 

\title{
Bilayer splitting in overdoped high $T_{c}$ cuprates.
} 

\author{S. E. Barnes and S. Maekawa }
\affiliation{Institute for Materials Research, Tohoku University, 
Sendai 980-8577, Japan}

\date{\today} 
\begin{abstract}
{Recent angle-resolved photoemission data for overdoped Bi2212 are explained. Of the peak-dip-hump structure, the peak corresponds the $\vec q =0$ component of a hole condensate which appears at $T_c$. The fluctuating part of this same condensate produces the hump. The bilayer splitting is large enough to produce a bonding hole and an electron antibonding quasiparticle Fermi surface.  Smaller bilayer splittings observed in some experiments
reflect the interaction of the peak structure with quasiparticle states near, but not at, the Fermi surface.  }
\end{abstract}


\maketitle

The existence of a bilayer splitting remains a controversial issue.  
There are now two separate claims\cite{1} to have observed via 
angle-resolved photoemission spectroscopy (ARPES)
such a splitting in modest to heavily overdoped Bi2212 while 
others\cite{2} have concluded that only a very small such splitting is 
possible for optimal or underdoped samples of the same material.  The 
question of a possible bilayer splitting cannot be separated from the 
recent debate about the nature of the Fermi surface.  Until recently 
the widely accepted Fermi surface was hole like\cite{3}, however there 
have been several reports\cite{4} that, for different ARPES photon 
energies, an electronlike Fermi surface is seen.  Very recently for a 
slightly overdoped Bi2212 it has been claimed\cite{5} that {\it 
both\/} electron {\it and\/} hole Fermi surfaces exist and that this 
reflects a modestly large bilayer splitting.  These most recent 
experiments would seem to be in conflict with the earlier observation 
of a bilayer splitting.  In the experiments \cite{1} of Feng  {\it et al.},  
on a heavily overdoped sample the antibonding band was observed to 
cross the Fermi surface near $[\pi, 0]$ while the existence of an 
electronlike antibonding Fermi surface for a smaller hole doping is 
at odds with this observation.

In the picture developed here, as proposed\cite{5} by Bogdanov {\it et al.}, 
there exists a bilayer splitting which is sufficiently large that the 
bonding band is hole, while the antibonding band is electronlike.  
That part of the hole like antibonding band seen\cite{1} by Feng {\it et al.}, 
which closes near $[\pi, 0]$, reflects, in fact, a near Fermi surface ``shadow"
resonance associated with an antibonding band which itself never 
crosses the Fermi surface.

For a bilayer the usual $t-J$ model becomes,
\begin{equation}
{\cal H} = - \sum_{ij\sigma\alpha\alpha^\prime} t_{i\alpha j\alpha^\prime}
[\hat c^\dagger_{i\sigma\alpha}\hat c^{}_{j\sigma\alpha^\prime} + {\rm H.c.}]
+ \sum_{ij\alpha\alpha^\prime}J_{i\alpha j\alpha^\prime} \vec S_{i \alpha}\cdot \vec S_{j \alpha}
\nonumber
\end{equation}
where $\hat c^\dagger_{i\sigma\alpha} = (1 - n_{i-\sigma\alpha} ) c^\dagger_{i\sigma\alpha}$; $n_{i\sigma\alpha}= c^\dagger_{i\sigma\alpha}c^{}_{i\sigma\alpha}$ and where $c^\dagger_{i\sigma\alpha}$ is for an electron of spin $\sigma$ site $i$ and plane $\alpha$. The in-plane interactions are $t$ and $J$ and near-neighbor. The interplane exchange $J_{\perp}$  is vertical, while the $t_{\perp}$ leads to a single-particle bilayer splitting $(t_{\perp}/4)(\cos k_x - \cos k_y)^2$. 

The present formalism signals a departure from the usual slave-boson method\cite{6,7}. Used here is a formalism based upon the Jordan-Wigner (JW) transformation\cite{8}. The system is mapped to a one dimensional path. A {\it single\/} spin fermion, $f^{\dagger}_{i}$ creates an up spin at {\it path\/} site $i$ when it acts on the down spin ferromagnetic vacuum $|\rangle_{-N/2}$. Here the total spin $S_{z} = - N/2$ and $N$ is the (even) number of sites. A hole is created by $b^{\dagger}_{i }$. A  hard-core constraint $ Q^{}_{i} = f^{\dagger}_{i} f^{}_{i} + b^{\dagger}_{i} b^{}_{i} \le 1$ is needed {\it only for finite hole doping\/} $x$. {\it Fictitious unit flux tubes\/} perform the JW transformation.  The tubes are reflected by {\it unitary\/} operators $u^{\dagger}_{i} = \exp[- i \int_{0}^{i-1} \vec a \cdot d \vec r]$ where $\vec a$ is a vector potential for tubes attached to each particle.\cite{8}  The raising operator $S_{i}^{+} = f^{\dagger}_{i}u^{}_{i}$ and destroys (in the sense $u^{}_{i} u^{\dagger}_{i} =1$) a flux tube before creating a spin particle, $S_{iz} = f^\dagger_{i}f^{}_{i } - {1\over 2} (1 - b^\dagger_{i } b^{}_{i})$.  The physical electron $\hat c^{\dagger}_{i  \uparrow } = f^{\dagger}_{i} b^{}_{i}$ while $\hat c^{\dagger}_{i \downarrow} = u^{\dagger}_{i} b^{}_{i}$ involves a flux tube.  

\begin{figure}
\centerline{\hskip -10pt \epsfig{file=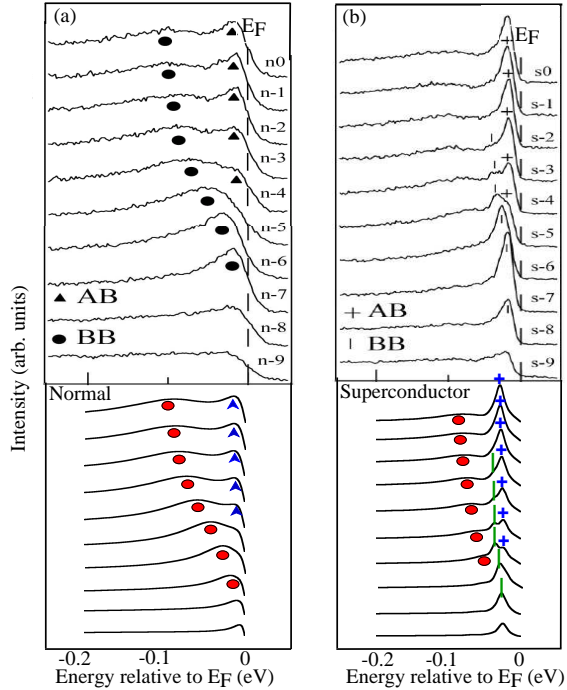,width=3.in}}
\vskip -5pt
\caption{
(a) Experiment\cite{1} (top) and theory (bottom) for the normal state.  
Theory corresponds to $\Im m{\cal G}_{a \vec k}$ with $s=35$ meV, $\sqrt{ab} = 
25$ meV $b = 0.65$, $d = 54$ meV and $e=35$ meV for $k_{y} = \pi$ and 
$k_{x} = 0\pi,0.025\pi, 0.05\pi, 0.075\pi \ldots 0.225\pi$ in 
descending order.  (b) The same for the superconducting state but with 
$s=6$ meV, $\sqrt{ab} = 31$ meV $b = 1.15$, $d = 50$ meV and $e=70$ meV. 
In both panels the strongly dispersive bonding band is indicated by a 
filled circle.  For the normal state the triangles indicate the 
position of the weakly dispersive residue of the antibonding band 
while in the superconductive state the $|$ and $+$ indicate the 
positions of the coherence peaks associated with the bonding and 
antibonding bands, respectively.  The rest of the parameters are given 
in the text.}
\label{f1}
\end{figure}

The standard mean-field slave-boson method introduces unphysical states because a certain constraint is obeyed only on average. The present development corrects this defect.  
This has important physical consequences.   Within the usual approach, for doping $x$, the effective spin hopping matrix element $ \sim (J+2xt)/2\sim 150$ meV which is sufficiently large to explain the strongly dispersive ``quasiparticle" features indicated by solid circles in the ARPES data and theory of Fig.~\ref{f1}.  In fact, this matrix element should be $J_e/2 = (J- 2xt)/2\sqrt{2}$ where the minus sign reflects the well-established fact that a small doping $x$ favors ferromagnetism. Near optimal doping $J_e$ is approximately zero. This $J_e$ is involved principally in the dispersion of the so called ``coherence peak," i.e., the ``+" and ``$|$" in Fig.~\ref{f1}. The ``quasiparticle" feature reflects a fluctuation part of the hole spectrum which is ``driven" by the constraint.  The physics of the $b$-particle hole Bose condensation is also changed when the constraint is enforced. Such a condensation needs the lowest energy level to be renormalized to the chemical potential and this results in a {\it gap equation\/} which determines $T_c$ in both the underdoped {\it and\/} overdoped regimes.  This paper introduces this interpretation of the ARPES data. In the process some rather subtle effects associated with the bilayer splitting are explained.

In analogy with the quantum Hall effect, the flux tubes reflected by the $u^{\dagger}_{i}$ are nontrivial. That $\hat c^{}_{i  \uparrow } = b^{\dagger}_{i}f^{}_{i} $ while $\hat  c^{}_{i \downarrow} = b^{\dagger}_{i}u^{}_{i} $ suggests that $f^{}_{i\alpha}$ and $u^{}_{i\alpha}$ are intimately related.  Important is the degeneracy of the vacuum. This is $|-N/2\rangle$, {\it however}, since $\cal H$ is rotationally invariant, the ferromagnetic state $|-N/2+1\rangle$ obtained by action of the {\it total spin\/} $S^+ $ on $|-N/2\rangle$ is equivalent. {\it Arbitrary\/} $N/2$-$f$-particle states are of the form $\psi_{N/2} |-N/2\rangle$ {\it or\/} $\phi_{N/2-1} |-N/2+1\rangle$ where $\psi_{N/2}$ and $\phi_{N/2-1}$ are suitable field operators which create respectively $N/2$ and $N/2-1$, $f$ particles. Consider $\hat c^{}_{i \downarrow } = b^{\dagger}_{i}u^{}_{i}$, it can be shown rigorously for large $N$ and $M\sim N/2$, $f$ particles,
\begin{eqnarray}
\langle -N/2+1 | \phi^\dagger_{M-1} b^{\dagger}_{i} u^{}_{i}\psi_{M} |-N/2\rangle \nonumber \\
= \langle -N/2 | \phi^\dagger_{M-1}b^{\dagger}_{i}f^{}_{i}\psi_{M} |-N/2\rangle.
\nonumber
%\label{deux}
\end{eqnarray}
Thus the {\it vacuum off-diagonal matrix elements\/} involving $u^{}_{i}$ are equal to {\it vacuum diagonal\/} matrix elements with this operator replaced by $f^{}_{i}$. Here
 $u^{}_{i}$  destroys a particle, as does $f^{}_{i}$, {\it but\/} the destroyed particle appears in the vacuum so $S_z$ is unchanged. A judicious choice of intermediate states is called for. With diagonal elements $u^{}_{i}$ is simply a  phase factor, its particle nature only being manifest when the intermediate states introduce off-diagonal elements.

Corresponding to a "flux state", in the absence of doping and treating the flux at the mean-field level, the spectrum of the $f$ particles is,
%\begin{equation}
${\cal E}_{ \vec k}= \sqrt{{J_e}^2{\gamma_{\vec k}}^2+{\Delta_0}^2{d_{\vec k}}^2}$,
%\nonumber
%\end{equation}
where $\gamma_{\vec k} = (\cos k_x + \cos k_y)$ while $d_{\vec k} = (\cos k_x - \cos k_y)$. A $T=0$, a relationship  $J_e = \Delta_0 = J/\sqrt{2}$ reflects the unitary nature of the $u^{}_{i}$.

Bose condensation involves the construction of a hole coherent state which satisfy the constraint. This is done by performing a SO(3) rotation. The space for a given site $i$ is spanned by the basis vectors $|\uparrow\rangle$, $|\downarrow\rangle$ and $|0\rangle$. The small rotation causes $|\sigma\rangle \to |\sigma\rangle + \theta_{\sigma i} |0\rangle$ while $|0\rangle \to |0\rangle - \sum_\sigma \theta_{\sigma i} |\sigma\rangle$. The rotation angle $\theta_i$ constitutes the wave function for the condensed holes. It is necessary to minimize the expectation value of ${\cal H}$ with respect to this rotation.  The result $(\partial {\cal H}/\partial \theta_i)=0$ is of the form of an equation of motion for $b^\dagger_i$ in which the energy has been equated to the chemical potential $\mu$; i.e., it amounts to the evident requirement that the lowest Bose level be renormalized to $\mu$. The hopping term for these bosons is $-\sum t_{ij}  u^\dagger_i b^{}_i b^\dagger_j u^{}_j$, and the resulting diagonal term in ${\cal H}$ is maximized in the negative sense when when $u_i \theta_{\downarrow i} = \theta$ a constant, while $u_i \theta_{\uparrow i} = \pm \theta$ where the sign alternates. Substituting this into $- \sum t_{ij} f^\dagger_i b^{}_i b^\dagger_j f^{}_j$ 
and making the same mean-field approximation as used for the $J$ term results in $ -\theta^2 (1/\sqrt{2})\sum t_{ij} f^\dagger_i f^{}_j$ which is without doubt ferromagnetic in nature, and $J_e = (J- 2xt)/\sqrt{2}$. A simple spin pair amplitude, e.g., $\langle f^\dagger_{\vec k } f^\dagger_{-\vec k }\rangle$, is forbidden by the constraint, however for half filling it is well known that the RVB state is reflected by a flux state amplitude $\langle f^\dagger_{\vec k } f^{}_{\vec k + \vec \pi}\rangle$, where $\vec \pi = (\pi, \pi)$. Since here the holes are subsumed into the spin particles, this result is extended to finite doping, i.e., with the above for $\theta_{\sigma i}$, it follows that the BCS pair amplitude $\langle \hat c^\dagger_{\vec k \uparrow} \hat c^\dagger_{-\vec k \downarrow}\rangle \approx \theta^2 \langle f^\dagger_{\vec k } f^{}_{\vec k + \vec \pi}\rangle$, i.e., the flux state plus {\it coherent\/} doping equals superconductivity.

The O(3) rotation of ${\cal H}$ introduces {\it new\/} $\theta^2$ terms of physical importance. Since $|00\rangle$, reflecting near-neighbor hole sites, mixes with $|\downarrow 0\rangle$, while $|\downarrow\downarrow\rangle$ mixes with $|0 \downarrow \rangle$, there is a $b$-particle pair term $- t {\theta}^2 u^\dagger_{ i} u^{}_{ j} b^{\dagger}_{i}b^{\dagger}_{ j} + {\rm H.c.}$ which results in a {\it charge gap\/}  $K_c \sim 4\theta^2 t$.   Rotating the physical electron operator $c^{\dagger}_{i\sigma }$ produces both $\theta_i f^{\dagger}_{i\sigma }$ {\it and\/} $\theta_i u^{\dagger}_{ i}f^\dagger_{i}f^{}_{i}$.  The former {\it coherent\/} term implies that  the condensate fraction is $c = \theta^2/2$ while the latter is corresponds to an {\it equal number\/} of constraint-driven {\it incoherent\/} fluctuating holes, this at zero temperature.

In other than one dimension,
the Bose levels are renormalized towards $\mu$ by the $-\sum t_{ij} f^\dagger_i b^{}_i b^\dagger_j f^{}_j$ term.  Using the usual renormalization group techniques this results in a {\it gap equation}. High-energy intermediate states are integrated out down to some cutoff $K(\mu)$. Since the intermediate states involve both spin and charge degrees of freedom there are two contributions to $K(\mu)=K_s + K_c$.  The charge pseudogap $K_c \sim 4\theta^2 t$ while $K_s \sim {\cal E}_{ \vec k=\pi,\pi/2}$ reflects the width of the spin band due to both dispersion and the spin BSC gap. The $T=0$ charge gap $K_c$  is the lowest-energy scale and $T_{c} \sim K_c$, i.e., $T_c \propto c$.  Spin pairing occurs at a higher mean-field temperature $T_s \sim J/ 2\sqrt{2}$ associated with the opening of a pseudogap. The scenario is supported 
the fact that $T_c \propto c$ and  evidence\cite{11} of a  pseudogap well above $T_{c}$, in tunneling for {\it overdoped\/} Bi${}_{2}$Sr${}_{2}$CuO${}_{6+\delta}$ and via Zeeman splitting\cite{11bis} in Bi${}_{2}$Sr${}_{2}Ca$CuO${}_{8+\delta}$.  With decreasing $\mu$, when condensation first occurs, $K(\mu)=K_s$ ($ K_c=0$). This point is {\it unstable\/} precisely because $J_e = (J- 2xt)/\sqrt{2} $, whence $K_s$ {\it decreases\/} with {\it increasing\/} $x$. In the underdoped region, when $\partial K_s/\partial x <0$, the system separates into hole-rich regions with $(J- 2xt)\sim 0$ and with hole-poor regions between. Doping occurs at constant $\mu$. In the overdoped region $\partial K_s/\partial x >0$ and implies that $K_c$ must decrease and with it the condensed fraction $c$ and $T_c$. Up to this point the holes have been accommodated within the rotated $f$ particles. Now $b$ particles must be added and this requires that $\mu$ decreases. This must increase $K(\mu)$ which {\it slows\/} the flow out of the condensate with increasing $x$. These $b$-particle holes are not particularly important for ARPES. They are hard-core bosons which must avoid the existing condensed bosons. This implies a curvature to their wave functions which pushes their energy to at least $\sim 4xt$, i.e., out of range of the dispersive features near the Fermi surface. Their presence {\it is\/} seen in $J_e = (J- 2xt)/\sqrt{2}$ since this is a mean-field result and all of the holes must be counted in the $x$.

The physical electron Green's function ${\cal G}^{}_{\vec k}$ describes the excitations of interest. Consider first a single plane. The operator$-$e.g., $\hat c^\dagger_{i\uparrow}\approx |b^{}_{i}|f^\dagger_{i}+  f^\dagger_{i} b^{\prime}_{i}$$-$contains a {\it condensate\/} part proportional to $|b^{}_{i}| = \theta/\sqrt{2}$ and a {\it noncondensate\/} part reflected by $b^{\prime}_{i}$. If within the Fourier transform of $-t\sum\hat c^\dagger_{i\sigma}\hat c^{}_{j\sigma} = -2t\sum \gamma_{\vec k}
\hat c^\dagger_{\vec k\sigma}\hat c^{}_{\vec k\sigma}$ the {\it single\/} matrix element $\gamma_{\vec k}$ is removed, then all coherent motion of {\it holes\/} with this wave-vector is suppressed and this decouples these two parts of ${\cal G}^{}_{\vec k}$. If ${\cal G}_{\vec k\ell}$ is the Green's function calculated without $\gamma_{\vec k}$, then trivially the full 
${\cal G}^{-1}_{\vec k} ={\cal G}^{-1}_{\vec k\ell} - (-2t)\gamma_{\vec k}$. A simple case would ignore both pairing and the noncondensate part, whence 
${\cal G}^{-1}_{\vec k\ell}=(\omega+is-J\gamma_{\vec k})/c$, leading to a ${\cal G}^{-1}_{\vec k}$ with poles at energy ${\cal E}_{\vec k} = (J-2ct)\gamma_{\vec k}/\sqrt{2}$, with strength $c$, and illustrating the contribution of the condensate to $J_e$.

When the bilayer coupling is included, eigenstates are simultaneous eigenvectors of $\cal H$ and a reflection operator $R$; i.e., solutions, with index $a=\pm$, are bonding or antibonding respectively.  Neutron data on the insulating compounds\cite{10} suggest that the inter-plane $J_{\perp}$ is vertical.  The bonding and antibonding $d$-wave {\it order parameters\/} must then be {\it identical\/} for symmetry reasons and $J_{\perp}$ is unimportant. With a bilayer slitting the condensate and noncondensate parts are decomposed by removing the matrix elements involving $\epsilon_{a \vec k} = - [2t\gamma_{\vec k} + 
a \frac{t_{\perp}}{4}(\cos k_x - \cos k_y )^{2}]$ (rather than $\gamma_{\vec k}$) and the Green's function becomes $
{\cal G}_{a \vec k}(\omega+is)
=
[(G_{a\vec k s}
+
 G_{f})/
(1
-
\epsilon_{a \vec k} 
 \left[
 G_{a \vec ks}
+
 G_{f}
 \right])$.

The full {\it condensate\/} contribution to ${\cal G}_{\vec k\ell}$ is easily seen to be
\begin{equation}
G_{a \vec k } \equiv 
\frac{c}{\omega+is - ((x-c)\epsilon_{a  \vec k} + J \gamma_{\vec k}) 
-{|\Delta d_{\vec k}|^{2}\over
\omega+is + (x\epsilon_{a - \vec k} + J \gamma_{-\vec k})}},
\label{huit}
\end{equation}
where included are the complications associated with {\it both\/} pairing and the bilayer splitting via $\epsilon_{a \vec k}$. As described above, the important {\it noncondensate\/} part corresponds, in fact, to the fluctuations of the condensate driven by the constraint. At this stage in the development of the theory this is  {\it modeled\/} by
\begin{equation}
G_{f} \equiv \frac{2 (x_c - c) }{(\omega + is - \Sigma_{f})},
\label{neuf}
\end{equation}
where $\Im m \Sigma_{f} = (a/\omega) + b(\omega - d) $ and where 
$\sqrt{ab} \sim K$.  Then 
$\Re e \Sigma_{f} = e\, {\rm sgn}(\omega)$ and $d$ also $\sim K$ 
result from shifts away from the Fermi surface due to the spin and charge gaps.  In order to simulate finite temperatures, $s$ is used to mimic thermal broadening, the condensed fraction $c$ is treated as a parameter, and  $x_c$ is the total number of bosons including those in the fluctuating part of the condensate {\it but\/} excluding the $b$ bosons mentioned above.   The ARPES and tunneling are simulated using the resulting ${\cal G}_{a \vec k}(\omega+is)$.

Normal state ARPES spectra are calculated with the scale $K \sim 
25$ meV with no holes condensed.  With the other parameters $x \approx 0.2$, 
$J=45$ meV, $2x_c t = 396$ meV and $2 x_c t_{\perp}=79$ meV, ${\cal G}_{a \vec k}(\omega+is)$  reproduces well the corresponding experimental spectrum of Feng et 
al.\cite{1}, Fig.~\ref{f1}a.  The {\it bonding band\/} lies at a
binding energy of $\sim 50-100$ meV and is quite broad.  The peak which 
lies within $\sim 20$ meV of the Fermi surface for smaller $k_{x}$ is a 
remnant of the {\it antibonding\/} band which lies {\it above\/} the 
Fermi surface.  The strength of this antibonding band has been 
artificially increased by a factor of $2$ in order to simulate the 
apparent experimental fact\cite{5} that under the conditions with 
which these spectra where taken, ARPES couples more strongly to the 
antibonding states.  The nature of the near-Fermi-surface peaks for 
$\vec k = [\pi,0]$ is not inconsistent with very recent data\cite{12} 
for overdoped Bi${}_{2}$Sr${}_{2}$CaCu${}_{2}$O${}_{8}$.  
Experimentally the ratio of the normal-state peak and hump intensities 
can be varied via the photon energy which changes the relative 
coupling to the bonding and antibonding bands.  Our simulations 
confirm {\it for this direction\/} that the hump and peak are 
predominantly of bonding and antibonding character, respectively.

\begin{figure}[t!]
	\vglue -0.3in
\centerline{\epsfig{file=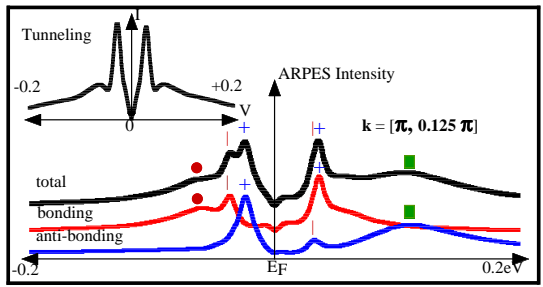,width=2.8in}  } 
\vspace{-3pt} 
\caption[toto]{The decomposition of theory into bonding 
and antibonding bands. The solid circle and square indicate the 
quasiparticle bands and the $|$ and $+$ the superconductive 
coherence peaks. The weight of the  antibonding band is 
multiplied by $2$ to simulate experiment. The inset shows theory for 
tunneling. }
\vglue -0.175in
\label{f2}
\end{figure}

The superconducting state implies hole condensation with a condensed fraction of $0.5$,  and with this coherence peaks develop near the Fermi surface.  The thermal broading reflected by  $s=6$ meV has been reduced. The reduction in the effect of thermal filling of the spin sector gap is reflected by an increased $K \sim 31$ meV.  The other parameters are the
same (see figure caption).  A quasiparticle {\it bonding\/} band, again at a binding
energy of $\sim 50-100$ meV, is identified in experiment spectra and
indicated in the theory by a solid circle.  The bonding and
antibonding coherence peaks identified by Feng {\it et al.}\cite{1} are
also indicated.   The decomposition into bonding and antibonding bands for
$\vec k = [\pi, 0.125\pi]$ is shown in Fig.~\ref{f2}.  The antibonding
band which lies above the Fermi surface ``pulls'' its coherence peak
towards the Fermi surface, leading to an apparent gap of only $\sim
20$ meV. On the other hand, the full gap $\sim 36$ meV is exhibited by
the coherence peak of the bonding band.

That the coherence peak splitting does {\it not\/} lead to a splitting 
of the tunneling spectrum, as it would in a simple BSC theory, is also illustrated by the inset in 
Fig.~\ref{f2}.  This spectrum was obtained with identical parameters 
as for the body of this figure by averaging over 100 points in the 
Brillioun zone.  It is interesting to note that the peak-dip-hump structure is 
reproduced with the dip at $\sim 60$ meV being consistent with typical 
tunneling data.\cite{13}

The value of $J=45$ meV reflects the gap via 
$\Delta = J/\sqrt{2}$.  This is small compared with $J=120$ meV 
deduced from a fit to the spin waves.\cite{10}  However, a quantum 
correction\cite{10} reduces the latter value to $J=100$ meV. The same correction 
reduces the antiferromagnetic order parameter by $\sim 60$\% and assuming a similar 
correction here increases the present $J$ to $ \sim 74$ meV. The difference is accounted for by a factor of $(1-x)$ which arises for the O(3) rotations and which was omitted in the above equations for clarity. 

The value of $2 x_c t_{\perp}=79$ meV is larger that the value $45$ meV 
deduced directly from experiment.\cite{1}  On the other hand, the 
larger bare value $t_{\perp} \approx 180$ meV agrees well with the local density approximation 
value.\cite{14}  That there is a strong renormalization of $t_{\perp}$ 
has been discussed elsewhere.\cite{15}

While the present JW formalism is intended for all doping levels, the present application has been to the overdoped regime. As described above, the present approach suggests that the underdoped region phase separates into hole-rich and hole-poor regions where the concentration in the hole-rich regions corresponds approximately to optimal doping.\cite{16}

This work was supported by a Grant-in-Ad for Scientific Research on
Priority Areas from the Ministry of Education, Science, Culture and
Technology of Japan, CREST. This work was mainly performed while SEB was on sabbatical leave from the
Physics Department, University of Miami, FL, U.S.A. and he wishes to
thank the members of IMR for their kind hospitality.    The authors wish to thank
D.~L.~Feng and Z.~X.~Shen for valuable communications about their
work.


\begin{thebibliography}{16}

\bibitem{1} D. L. Feng, N. P. Armitage, D. H. Lu, A. Damascelli, J. P.
Hu, P. Bogdanov, A. Lanzara, F. Ronning, K. M. Shen,
H. Eisaki, C. Kim, Z.-X. Shen, J. -I. Shimoyam, and K. Kishio,  Phys.  Rev.  Lett.  {\bf 86}, 5550 
(2001); Y.-D Chuang, A. D. Gromko, A. Fedorov, Y. Aiura, K. Oka,
Y. Ando, H. Eisaki, S. I. Uchida, and D. S. Dessau, Phys.  Rev.  Lett.  {\bf 87} 117002 (2001).

\bibitem{2} J. Mesot , M. Randeria, M. R. Norman, A. Kaminski, H. M.
Fretwell, J. C. Campuzano, H. Ding, T. Takeuchi, T. Sato,
T. Yokoya, T. Takahashi, I. Chong, T. Terashima, M. Takano, T. Mochiku, and K. Kadowaki,  Phys.  Rev.  B{\bf 63}, 224516 (2001).

\bibitem{3} P.~Aebi, J. Osterwalder, P. Schwaller, L. Schlapbach, M. Shimoda,
T. Mochiku, and K. Kadowaki, Phys.  Rev.  Lett.  {\bf 72} 2757 (1994);
Jian~Ma, C. Quitmann, and R. J. Kelley, P. Alm\'eras, H. Berger, and G. Margaritondo, and M. Onellion, Phys.  Rev.  B{\bf 51} 3832 (1995); A.~G.~Loeser, Z.-X. Shen, D. S. Dessau, D. S. Marshall, C. H. Park, P. Fournier, and A. Kapitulnik, Science {\bf 273} 325 (1996); 
P.~J.~White, Z.-X. Shen, C. Kim, J. M. Harris, A. G. Loeser,
P. Fournier, and A. Kapitulnik, Phys.  Rev. 
B{\bf 54} 15669 (1996); 
H. Ding, A. F. Bellman, J. C. Campuzano, M. Randeria,
M. R. Norman, T. Yokoya, T. Takahashi, H. Katayama-
Yoshida, T. Mochiku, K. Kadowaki, G. Jennings, and G. P.
Brivio, Phys.  Rev.  Lett.  {\bf 76}
1533 (1996); 
M.~R.~Norman, H. Ding, M. Randeria, J. C. Campuzano,
T. Yokoya, T. Takeuchi, T. Takahashi, T. Mochiku,
K. Kadowaki, P. Guptasarma, and D. G. Hinks, Nature {\bf 392} 157 (1998); N.~L.~Saini, J. Avila, A. Bianconi, A. Lanzara, M. C. Asensio,
S. Tajima, G. D. Gu, and N. Koshizuka, Phys.  Rev.  Lett.  {\bf 79} 3467 (1997).

\bibitem{4} Y.-D.~Chuang, A. D. Gromko, D. S. Dessau, Y. Aiura, Y. Yamaguchi,
K. Oka, A. J. Arko, J. Joyce, H. Eisaki, S. I.
Uchida, S. I. Uchida, K. Nakamara, and Yoichi Ando, Phys. Rev. Lett. {\bf 83} 3717 (1999); 
D.~L.~Feng, W. J. Zheng, K. M. Shen, D. H. Lu, F. Ronning,
J. Shimoyama, K. Kishio, G. Gu, D. V. der Marel, and
Z.-X. Shen, cond-mat/9908056 (unpublished); 
A.~D.~Gromko, Y.-D. Chuang, D. S. Dessau, K. Nakamura,
and Y. Ando, cond-mat/0003017 (unpublished).

\bibitem{5} P.~V.~Bogdanov, A. Lanzara, X. J. Zhou, S. A. Kellar, D. L. Feng, E. D. Lu, H. Eisaki, J.-I. Shimoyama, K. Kishio, Z. Hussain, and Z.-X. Shen
Phys. Rev. B 64, 180505 (2001).

\bibitem{6} S.~E.~Barnes,  J. Phys. F{\bf 6} 115, 1375 (1976);
J. Phys. F{\bf 7} 2637 (1976);  Adv. Phys. {\bf 30} 801 (1980).



\bibitem{7} G. Baskaran, Z. Zuo and P. W. Anderson, Solid State Commun. 
{\bf 63}, 973 (1987).

\bibitem{8} S.~E.~Barnes and S.~Maekawa, J. Phys. Condens. Matter {\bf 14}, L19, (2002); cond-mat/0111204 (unpublished).



\bibitem{11} M. Kugler, \O. Fischer, C. Renner S. Ono and Y. Ando,
Phys.  Rev.  Lett.  {\bf 86}, 4911 (2001).

\bibitem{11bis} T. Shibauchi, L. Krusin-Elbaum, Ming Li, M. P. Maley, and P. H. Kes,
Phys.  Rev.  Lett.  {\bf 86}, 5763 (2001).

\bibitem{10} D. Reznik, P. Bourges, H.F. Fong, L.P. Regnault, J. Bossy, C. Vettier, D.L. Milius, I.A. Aksay and  B. Keimer, Phys.  Rev. B{\b 53} R14741 (1996).

\bibitem{12} A. A. Kordyuk, S. V. Borisenko, T. K. Kim, K. Nenkov, M. 
Knupfer, M. S. Golden, J. Fink, H. Berger and R. Follath, Phys.  Rev.  Lett.  {\bf 89} 077003 (1997).

\bibitem{13} see e.g., S. H. Pan, E. W. Hudson, A. K. Gupta, K.-W. Ng, H. Eisaki, S. Uchida, and J. C. Davis, Phys. Rev. Lett. {\bf 85}, 1536 (2000).

\bibitem{14} S. Chakravaty, A. Subdo, P.W. Anderson, S. Strong,  Science {\bf261}, 337 (1993); 
O. K. Anderson, A.I. Liechenstein, O Jepson, and F. Paulson, J. Phys. Chem Solids {\bf 12}, 1573 (1995).
	   
\bibitem{15} A. I. Liechtenstein, O. Gunnarsson, O. K. Andersen, and
R. M. Martin, Phys. Rev.
B{\bf 54}, 12505 (1996); R. Eder, Y. Ohta, and S. Maekawa,  Phys.  Rev. 
B{\b 51} 3265 (1995).

\bibitem{16} Experiment: C. Howald, P. Fournier and A. Kapitulnik, Phys.  Rev. B{\b 64} 100504 (2001); K. M. Lang, V. Madhavan, J.E. Hoffman, E.W. Hudson, H. Eisaki, S. Uchida, and J.C. Davis, Nature {\bf 415}, 412 (2002), manifests such a localization.

\end{thebibliography}
\end{document}